\documentclass[conference,twocolumn,10pt,a4paper]{IEEEtran} 

\pdfoutput=1
\usepackage[cmex10]{amsmath}
\usepackage{amsfonts,amssymb}
\usepackage{verbatim}

\usepackage{graphicx}
\graphicspath{{graphics/}}

\usepackage{cite}

\usepackage{array}

\usepackage{mdwmath}
\usepackage{mdwtab}

\newcommand{\numPop}{n_\textrm{T}}
\newcommand{\numCoop}{n_\textrm{c}}
\newcommand{\numSelf}{n_\textrm{s}}

\newcommand{\food}{R}
\newcommand{\eff}{\epsilon}
\newcommand{\effCoop}{\epsilon_\textrm{c}}
\newcommand{\effSelf}{\epsilon_\textrm{s}}
\newcommand{\effSelfPos}{\epsilon_\textrm{s}^+}
\newcommand{\effSelfNeg}{\epsilon_\textrm{s}^-}
\newcommand{\coopMult}{\alpha_\textrm{c}}
\newcommand{\selfPen}[1]{\beta_\textrm{s}^{#1}}
\newcommand{\selfPenPos}{\selfPen{+}}
\newcommand{\selfPenNeg}{\selfPen{-}}

\newcommand{\costCoop}{\gamma_\textrm{c}}
\newcommand{\costSelf}{\gamma_\textrm{s}}

\newcommand{\rev}{A}

\newcommand{\payoff}{U}

\newcommand{\payoffCoop}{U_\textrm{c}}
\newcommand{\payoffSelf}{U_\textrm{s}}
\newcommand{\payoffTotal}{U_\textrm{tot}}

\newcommand{\numPopEst}[1]{\hat{n}_{\textrm{T}#1}}
\newcommand{\numCoopEst}[1]{\hat{n}_{\textrm{c}#1}}
\newcommand{\probDetect}[1]{p_{\textrm{d}#1}}
\newcommand{\probFA}{p_\textrm{FA}}
\newcommand{\detectMult}{\eta}
\newcommand{\detectMultCoop}{\eta_\textrm{c}}
\newcommand{\detectMultSelf}{\eta_\textrm{s}}
\newcommand{\dist}[1]{d_{#1}}
\newcommand{\boxWidth}{h}

\newcommand{\EXP}[1]{\exp\left(#1\right)}

\begin{document}
\bibliographystyle{IEEEtran}
%
\title{Effect of Local Population Uncertainty on Cooperation in Bacteria}

\IEEEspecialpapernotice{(Invited Paper)}

\author{\IEEEauthorblockN{Adam Noel\IEEEauthorrefmark{1}, Yuting Fang\IEEEauthorrefmark{2}, Nan Yang\IEEEauthorrefmark{2}, Dimitrios Makrakis\IEEEauthorrefmark{1}, and Andrew W. Eckford\IEEEauthorrefmark{3}}
	\IEEEauthorblockA{\IEEEauthorrefmark{1}School of Electrical Engineering and Computer Science,
		University of Ottawa, Ottawa, ON, Canada}
	\IEEEauthorblockA{\IEEEauthorrefmark{2}Research School of Engineering,
	Australian National University, Canberra, ACT, Australia}
	\IEEEauthorblockA{\IEEEauthorrefmark{3}Dept. of EECS,
		York University, Toronto, ON, Canada}}


\maketitle

\begin{abstract}
	Bacteria populations rely on mechanisms such as quorum sensing to coordinate complex tasks that cannot be achieved by a single bacterium. Quorum sensing is used to measure the local bacteria population density, and it controls cooperation by ensuring that a bacterium only commits the resources for cooperation when it expects its neighbors to reciprocate. This paper proposes a simple model for sharing a resource in a bacterial environment, where knowledge of the population influences each bacterium's behavior. Game theory is used to model the behavioral dynamics, where the net payoff (i.e., utility) for each bacterium is a function of its current behavior and that of the other bacteria. The game is first evaluated with perfect knowledge of the population. Then, the unreliability of diffusion introduces uncertainty in the local population estimate and changes the perceived payoffs. The results demonstrate the sensitivity to the system parameters and how population uncertainty can overcome a lack of explicit coordination.
\end{abstract}

\section{Introduction}

Game theory, a formal mathematical method for modeling interactive behavior, has a long history in fields as diverse as economics, sociology, political science, and wireless communications \cite{Srivastava2005}. Mathematical games permit the theoretical study of complex behavior such as competition and cooperation. These games work the same way as, for example, a game of chess: players have certain allowed actions (or ``moves''), and the moves of all the players interact, resulting in specified rewards or losses for the players. 

Game theory has also long been useful in biology, as it provides a mathematical framework to show how species in competition can arrive at interdependent optimal solutions for their evolutionary fitness \cite{Schuster2008}. In this paper, we focus on studying the ``game'' between cooperating and/or competing bacteria. Unlike human players, bacteria are not conscious decision makers. However, they have evolved biochemical mechanisms that can activate discrete actions in response to a given stimulus, e.g., the run-and-tumble motion of {\em E. Coli} chemotaxis, which enables bacteria to navigate nutrient gradients; see \cite{Berg2004}. These discrete actions can be modeled as a game with rewards and losses, such as finding a food source or starving. Game theory provides a framework for evaluating the overall fitness of these mechanisms, which may lead to insights about how these mechanisms evolved into their present form and how they could be controlled.

This paper introduces a simple model for a game where a population consumes a common fixed resource. The model is inspired by studies of cooperation and competition in bacteria, where many complex game dynamics have been identified; see \cite{Velicer2003,Schuster2008,Hummert2014}. Every member of the population either competes or cooperates to harvest the resource, and the efficiency with which one consumes the resource depends both on its behavior and that of the other members of the population. The details of the model are chosen for simplicity while also reflecting practical behavior in microbial systems. Specifically, our influences include quorum sensing \cite{Atkinson2009}, population saturation \cite{Lambert2014}, selfish defection (i.e., cheating) \cite{Velicer2003}, and cheater punishment mechanisms \cite{Travisano2004}. Quorum sensing, where the local population is estimated to control cooperation, is a well-studied example of cooperation via signaling in bacteria; see \cite{Brown2001}. In our model, we account for imperfect population estimation due to diffusion noise. This uncertainty can actually overcome a lack of explicit coordination. The main contributions of this paper are the following:
\begin{enumerate}
	\item We define a payoff model for consuming a common resource that is suitable for application to bacteria.
	\item We describe a dynamic game where the bacteria are players and they decide whether to cooperate based on knowledge of the current population. Noisy detection is inspired by quorum sensing, where signaling success depends on the distance between the players.
	\item We play the game numerically to demonstrate how the model supports experimental observations and to observe that population uncertainty can lead to more cooperation and better payoffs than perfect knowledge.
\end{enumerate}

Other related work includes the following. A recent paper on the fundamental limits of quorum sensing with asymptotically large populations is \cite{Michelusi2017}. In \cite{Canzian2014}, game-theoretic models were introduced to show how individual links in a bacterial network could form. A game-theoretic model was given in \cite{Koca2017} to show the effects of cooperation and uncertainty on communication efficiency within a nanoscale network.

\section{System Model}
\label{sec_model}

In this section, we present a model for resource (e.g., food) access and sharing within a single population. We frame the model from the perspective of bacteria, but it could also be applied to other populations that share a common resource. Our goal in this work is not quantitative precision, but instead to have a simple analytical model whose output is qualitatively consistent with real-world observations of bacteria.

We consider a population with $\numPop$ members in an environment with a total resource availability (or availability rate) $\food$. At any given moment, there are $\numCoop$ cooperating members ($1 \le \numCoop \le \numPop$) and $\numSelf = \numPop-\numCoop$ selfish (i.e., competing) members. Every member pays a cost for its behavior, which represents the consumption of energy needed to act. Cooperating and selfish members pay $\costCoop$ and $\costSelf$, respectively. We assume that cooperation is more expensive than selfishness, i.e., $\costCoop>\costSelf$, as supported by models in \cite{Broom2013,Canzian2014,Lambert2014}.

All members of the population, whether they are cooperating or selfish, divide the available resource $\food$ among themselves. We impose that each type of member accesses the resource with a different efficiency $\eff$, where $0 \le \eff \le 1$, and the value depends on the population distribution. We define the efficiency as follows. We assume that cooperating members work together to improve their efficiency, such that they become more efficient when more cooperators work together. This is the underlying motivation for quorum sensing; see \cite{Atkinson2009}. However, we also impose diminishing returns as more cooperators join, since populations saturate as they consume all of the resources in an environment, e.g., see \cite{Lambert2014}. A simple model that satisfies these assumptions is exponential, and we write the cooperating efficiency $\effCoop$ as
\begin{equation}
\label{eqn_coop_efficiency}
\effCoop = 1-\EXP{-\coopMult\numCoop},
\end{equation}
where $\coopMult$ is a cooperation multiplier and $\coopMult > 0$.

For the selfish members, we consider two different efficiency models. If the selfish members are able to take advantage of the cooperators, e.g., by acquiring better access to the resource, then their efficiency improves with the presence of cooperators. This behavior is known as \emph{defection}, which underlies the classical Prisoner's Dilemma game (see \cite[Ch.~4]{Broom2013}) and has also been observed within bacteria populations including \emph{E. coli} and \emph{Myxococcus}; see \cite{Velicer2003}. In this case, a simple model for selfish efficiency $\effSelfPos$ is
\begin{equation}
\label{eqn_self_efficiency_pos}
\effSelfPos = \frac{\effCoop}{\selfPenPos},
\end{equation}
where $\selfPenPos$ is a selfishness penalty. We assume that $\selfPenPos>1$, but this does not prevent a selfish member from having a higher net payoff, since cooperation has a higher cost.

It is also possible that non-cooperative members of a population are prevented from accessing the benefits of cooperation, i.e., the selfish members are \emph{punished}. For example, cooperators may generate toxins that are only harmful to selfish members; see \cite{Travisano2004} for details and other possible punishment mechanisms for microbes. We model this by decreasing the selfish efficiency $\effSelfNeg$ as cooperation increases and write
\begin{equation}
\label{eqn_self_efficiency_neg}
\effSelfNeg = \frac{1-\effCoop}{\selfPenNeg},
\end{equation}
where again $\selfPenNeg$ is a selfishness penalty and $\selfPenNeg > 1$.

Given a member's efficiency $\eff$, we define its revenue $\rev$ as
\begin{equation}
\label{eqn_revenue}
\rev = \frac{\eff}{\numPop}\food.
\end{equation}

The net payoff $\payoff$, which we will hereafter refer to as the payoff, is the difference between the revenue and the cost. Thus, a cooperating member's payoff $\payoffCoop$ is
\begin{equation}
\label{eqn_coop_payoff}
\payoffCoop = \frac{\effCoop}{\numPop}\food - \costCoop = \frac{1-\EXP{-\coopMult\numCoop}}{\numPop}\food - \costCoop,
\end{equation}
and a selfish member's payoff $\payoffSelf$ is
\begin{equation}
\label{eqn_self_payoff}
\payoffSelf = \frac{\effSelf}{\numPop}\food - \costSelf =
\left\{
\begin{array}{cl}
\frac{\effCoop}{\selfPenPos\numPop}\food - \costSelf, & \textrm{if gain from cooperation} \\
\frac{1-\effCoop}{\selfPenNeg\numPop}\food - \costSelf, & \textrm{if lose from cooperation}.
\end{array}
\right. 
\end{equation}

The total payoff $\payoffTotal$ for the entire population is then
\begin{equation}
\label{eqn_payoff_total}
\payoffTotal = \numCoop\left(\frac{\effCoop}{\numPop}\food - \costCoop\right) +
\numSelf\left(\frac{\effSelf}{\numPop}\food - \costSelf\right).
\end{equation}

\begin{figure}[!tb]
	\centering
	\includegraphics[width=\linewidth]{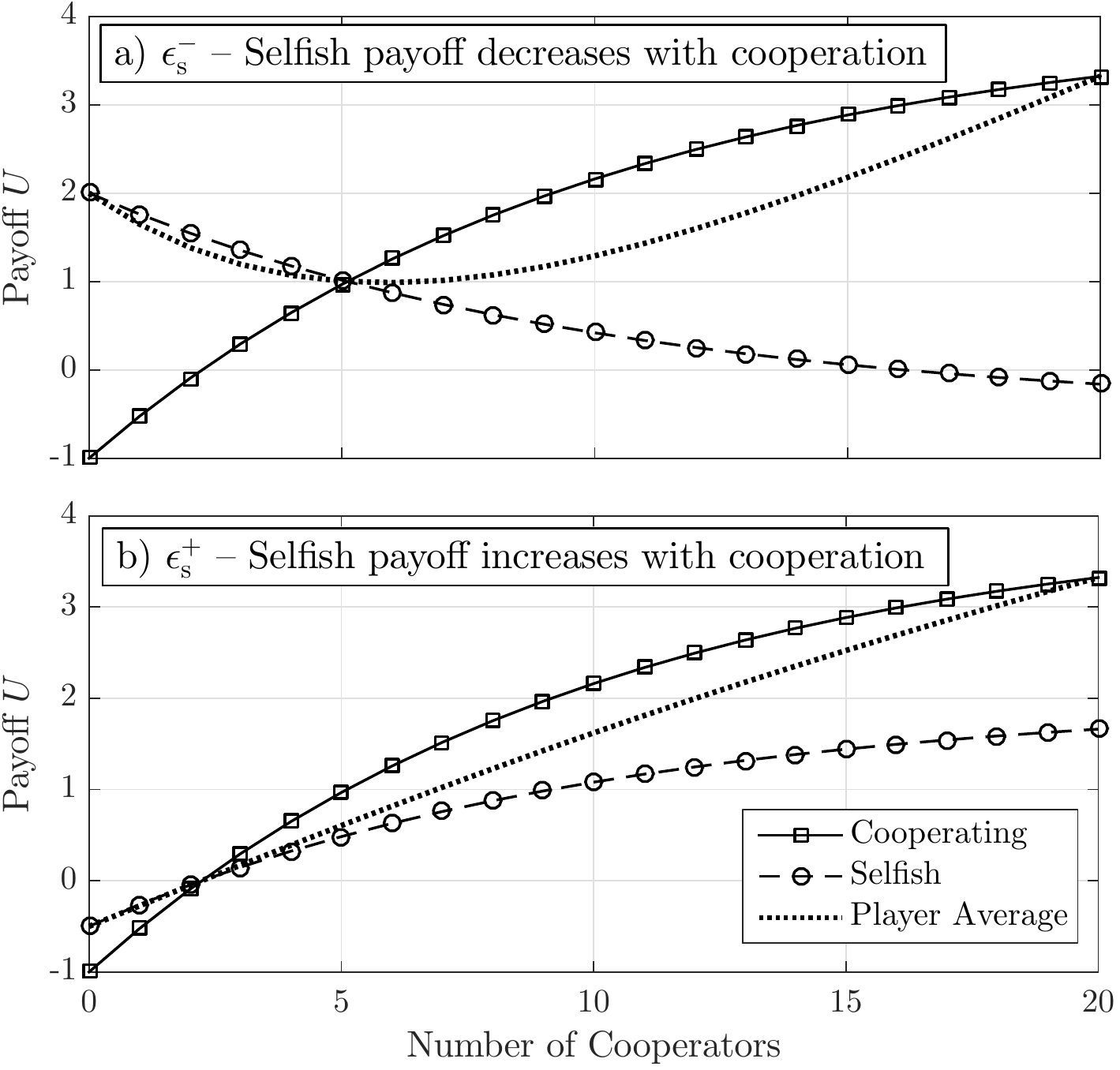}
	\caption{Member payoff as a function of the number of cooperators in a population of size $\numPop=20$ members. The remaining parameters are described in Table~\ref{table_param}.}
	\label{fig_payoff_sample}
\end{figure}

As an example of member payoffs, consider the system whose parameters are listed in Table~\ref{table_param} and which are the system defaults for the remainder of this paper. For this system, we vary the number of cooperators $\numCoop$ and plot the corresponding payoffs for each cooperator, selfish member, and average member in Fig.~\ref{fig_payoff_sample}. Our interest is in the difference between the payoff curves and how each curve varies with $\numCoop$, but \emph{not} on the absolute values of the payoffs (including whether they are negative). We consider the cases where selfish payoffs decrease and increase with increasing cooperation in Fig.~\ref{fig_payoff_sample}a) and b), respectively. Both cases have ranges where there is a higher payoff to cooperate or a higher payoff to be selfish. In the following section, we present a game where the population members are players and they have opportunities to change their behavior to get a higher payoff.

\begin{table}[!tb]
	\centering
	\caption{Default system payoff parameters.}
	
	{\renewcommand{\arraystretch}{1.4}
		\begin{tabular}{l|c||c}
			\hline
			\bfseries Parameter & \bfseries Symbol & \bfseries Value \\ \hline \hline
			Resource Availability & $\food$
			& 100 \\ \hline
			Total Population & $\numPop$
			& 20 \\ \hline
			Cooperation Cost & $\costCoop$
			& 1 \\ \hline
			Selfish Cost & $\costSelf$
			& 0.5 \\ \hline
			Cooperation Multiplier
			& $\coopMult$
			& 0.1 \\ \hline
			Selfishness Penalty & $\selfPenPos,\selfPenNeg$
			& 2 \\ \hline
		\end{tabular}
	}
	\label{table_param}
\end{table}

\section{Bacteria Cooperation Game}
\label{sec_game}

In this section, we present the bacteria game dynamics that we will simulate in Section~\ref{sec_results}. The members of the population are hereafter referred to as the players, and each player's strategy may change over time. We do not consider bacteria to be rational decision-makers, as is typical in classical game theory. Nevertheless, game theoretic analysis is suitable, as emphasized in \cite{Schuster2008,Velicer2003}, because bacteria do change their behavior in response to external signals. From this perspective, each player's strategy is whether it cooperates or acts selfishly.

We model a dynamic game as follows. Every player has some fixed initial behavior, i.e., to be selfish or to cooperate. Then, every player has the opportunity to maintain its current behavior or switch to the alternate behavior. These assessments are made simultaneously, independently of the other players, and ``selfishly'', i.e., a player will only cooperate if there is a perceived individual benefit, based on the information available. The game reaches a Nash equilibrium when no member can benefit by changing behaviors (see \cite[Ch.~3]{Broom2013}), i.e., when the population distribution converges. In the following, we consider perfect knowledge as an ideal case and then imperfect knowledge due to the randomness of diffusive signaling.

\subsection{Perfect Population Knowledge}

In the perfect (ideal) case, every player knows the current population size $\numPop$ and the number of cooperators $\numCoop$. If a player is currently cooperating, then it compares its current payoff $\payoffCoop$, found via (\ref{eqn_coop_payoff}), with the potential selfish payoff $\payoffSelf$, found via (\ref{eqn_self_payoff}) and where the number of cooperators when finding $\effCoop$ is decremented by 1. Analogously, if a player is currently selfish, then it compares its current payoff $\payoffSelf$, found via (\ref{eqn_self_payoff}), with the potential cooperating payoff $\payoffCoop$, found via (\ref{eqn_coop_payoff}) and where the number of cooperators when finding $\effCoop$ is incremented by 1. When no player can benefit by changing its behavior, then a Nash equilibrium is achieved.

\begin{figure}[!tb]
	\centering
	\includegraphics[width=\linewidth]{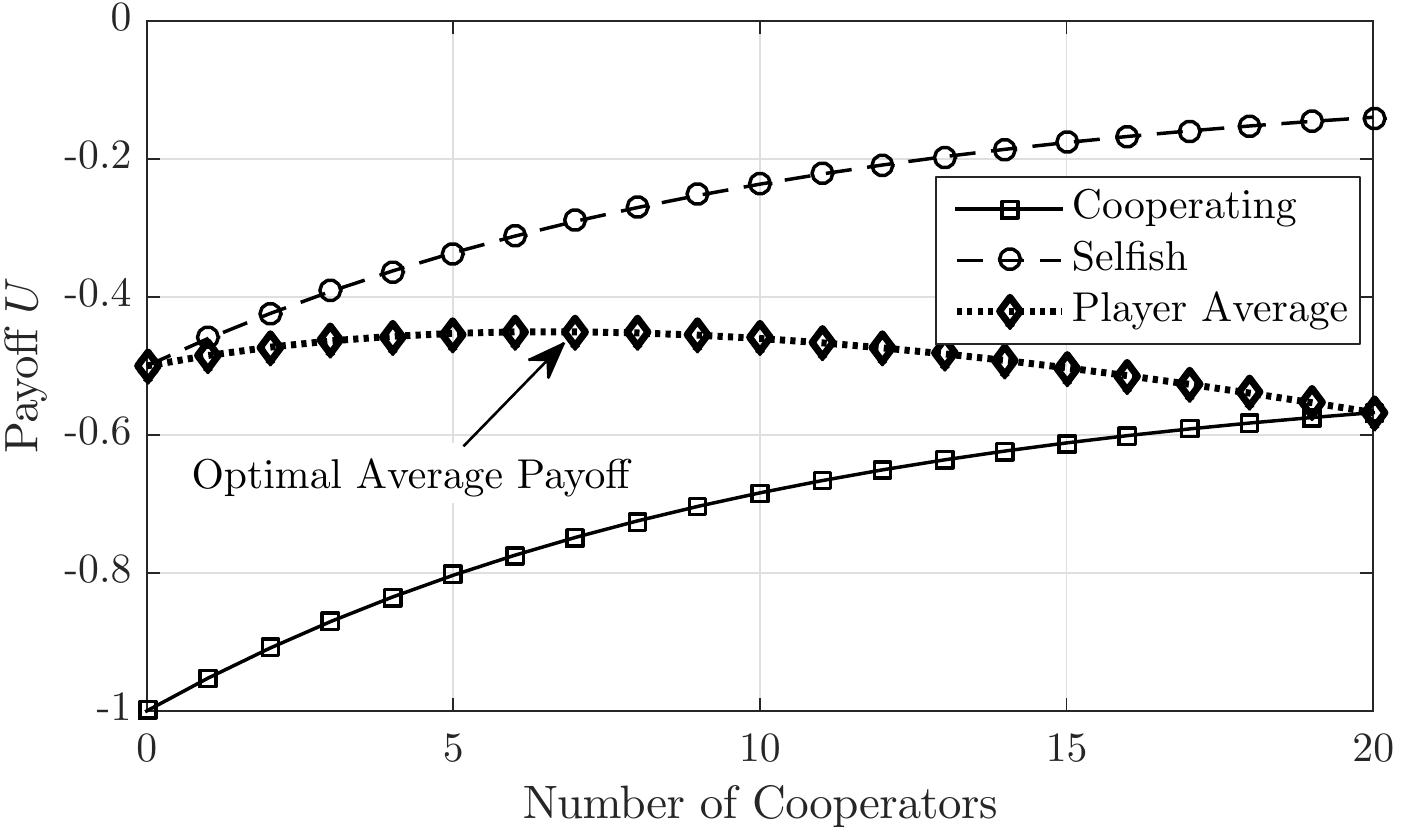}
	\caption{Player payoff as a function of the number of cooperators. Payoff for each cooperating player, selfish player, and average player is plotted separately. The selfish efficiency increases with the number of cooperators. The total resource level and the penalty for selfish behavior are reduced to $\food=20$ and $\selfPenPos=1.2$. Otherwise, the system parameters are the same as those specified in Table~\ref{table_param}.}
	\label{fig_payoff_optimal_non_ne}
\end{figure}

Both Figs.~\ref{fig_payoff_sample}a) and b) show that there are two Nash equilibria (zero cooperators or all cooperators), even though there is clearly one optimal strategy, since all players cooperating leads to the largest average payoff. The equilibria can be verified by confirming that, at equilibrium, a player's payoff will decrease if it switches strategies. We also note that the optimal solution is not always a Nash equilibrium. For example, consider Fig.~\ref{fig_payoff_optimal_non_ne}, where we decrease the total resource availability $\food$ and the selfishness penalty $\selfPenPos$. The highest average (i.e., mean) payoff is when there are 7 cooperators, but this will not be maintained in a game because the cooperating players can always increase their individual payoffs by becoming selfish.

\subsection{Noisy Population Detection}

In the noisy case, each player has to estimate the size of the population and the number of cooperators. We assume that the other payoff parameters are known. To simplify our model, we propose a proxy for bacteria quorum sensing. We define a probability of detection that increases as the distance between two players decreases. This is appropriate because the energy in a diffusing pulse of molecules is inversely proportional to the distance traveled, and a pulse with more energy is easier to detect; see \cite{Llatser2013}. We write the probability of the $i$th player detecting the presence of the $j$th player, $\probDetect{,i,j}$, as
\begin{equation}
\label{eqn_prob_detect}
\probDetect{,i,j} = 1 - \EXP{-\frac{\detectMult}{\dist{i,j}}},
\end{equation}
where $\dist{i,j}$ is the distance between the two players and $\detectMult$ is a detection scaling factor. We assume that cooperative players are easier to detect (e.g., by releasing more molecules), since they are motivated to attract other cooperators, so their detection scaling factor is larger than that of selfish players, i.e., $\detectMultCoop > \detectMultSelf$.

The $i$th player's estimate of the population size, $\numPopEst{,i}$, is then
\begin{equation}
\label{eqn_pop_est}
\numPopEst{,i} = 1 + \sum_{j \ne i} \mathcal{B}(\probDetect{,i,j}),
\end{equation}
where $\mathcal{B}(\probDetect{,i,j})$ is a binary number drawn from a Bernoulli distribution with probability $\probDetect{,i,j}$, and $\probDetect{,i,i}=1$. To estimate the number of cooperating players, $\numCoopEst{,i}$, the $i$th player classifies every other player that they detected using (\ref{eqn_pop_est}). For simplicity, we assume that classifying cooperators is as unreliable as their detection, such that they are classified with probability $\probDetect{,i,j}$ (i.e., they are misclassified as selfish with probability $1-\probDetect{,i,j}$). It is less intuitive for distant selfish players to be misclassified as cooperators with a high probability, so we mis-classify detected selfish players with a constant ``false alarm'' probability $\probFA$.

Given each player's estimates $\numPopEst{,i}$ and $\numCoopEst{,i}$, the game proceeds analogously to the perfect knowledge case. The current and prospective payoffs are estimated using (\ref{eqn_coop_payoff}) and (\ref{eqn_self_payoff}), where $\numPop$ and $\numCoop$ are replaced with their corresponding estimates. However, as we will demonstrate in the following section, the progress of the noisy detection game can be quite distinct from the perfect knowledge game, such that the uncertainty can overcome a lack of explicit coordination.

\section{Numerical Results}
\label{sec_results}

In this section, we execute examples of the bacteria cooperation game. Unless otherwise noted, we consider the payoff parameters listed in Table~\ref{table_param}, and we use $\selfPenNeg$ to evaluate the selfish efficiency $\effSelfNeg$. Every game is run for 100 rounds, and in each round every player has the opportunity to switch behaviors. In the perfect knowledge game, we do not need to account for the locations of the players. In the noisy detection game, we assume independent random motion and place the players as follows. We define the environment as a cube of width $\boxWidth$. In each round, a uniformly-distributed location is generated within the cube for every player. From these locations, the distance $\dist{i,j}$ terms for the evaluation of (\ref{eqn_prob_detect}) can be found. The system width $\boxWidth$ and the other detection parameters are as specified in Table~\ref{table_detection}.

\begin{table}[!tb]
	\centering
	\caption{Default detection parameters.}
	
	{\renewcommand{\arraystretch}{1.4}
		\begin{tabular}{l|c||c}
			\hline
			\bfseries Parameter & \bfseries Symbol & \bfseries Value \\ \hline \hline
			System Width & $\boxWidth$
			& $10\,\mu$m \\ \hline
			Cooperation Detection Factor & $\detectMultCoop$
			& $20\times10^{-6}$ \\ \hline
			Selfish Detection Factor & $\detectMultSelf$
			& $5\times10^{-6}$ \\ \hline
			False Alarm Probability & $\probFA$
			& $0.2$ \\ \hline
		\end{tabular}
	}
	\label{table_detection}
\end{table}

\begin{figure}[!tb]
	\centering
	\includegraphics[width=\linewidth]{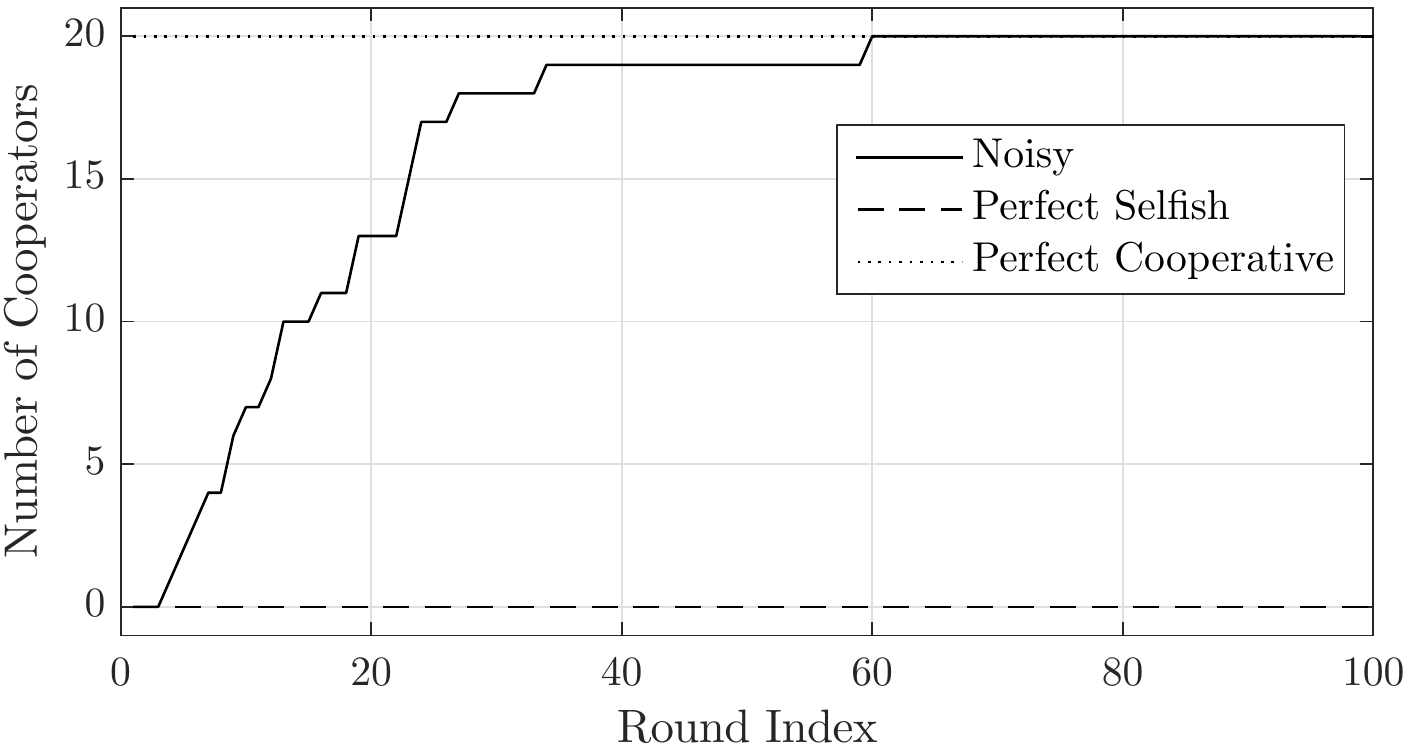}
	\caption{Progression of a game where the number of cooperators is measured in each round. A noisy game where the players are initially selfish is compared with games where players have perfect information and begin with either all players cooperative or all players selfish.}
	\label{fig_game_noisy_vs_perfect}
\end{figure}

In Fig.~\ref{fig_game_noisy_vs_perfect}, we consider the progression of a game with the default system parameters. The perfect knowledge game is run with all players initially selfish and again with all players initially cooperating. As we discussed in Section~\ref{sec_game}, these two cases are Nash equilibria, so no player modifies its behavior and the population remains stable for the entire duration of the game. We also consider the noisy detection game where all players are initially selfish. In this game, the imperfect knowledge leads to cases of misclassification, such that the players expect payoffs from cooperation that are larger than what is currently possible, but after 60 rounds the population has been ``pushed'' towards the equilibrium where all of the players are cooperating and the total payoff is maximized (as shown in Fig.~\ref{fig_payoff_sample}a)). Thus, noisy signaling is shown to promote optimal behavior among the bacteria, even though the players are not explicitly coordinating their strategies.

In the remaining figures, we focus on the noisy detection game and its sensitivity to individual system parameters. We vary one parameter at a time and for each value we run at least 10 games of 100 rounds each. We plot the mean number of cooperating players $\numCoop$ and mean total payoff $\payoffTotal$ over the \emph{entire duration} of all of the games. We separately consider the cases where all players are initially cooperating and where all players are initially selfish. In many of the cases considered, the population reaches an equilibrium where either all or none of the players cooperate, independent of the initial state.

\begin{figure}[!tb]
	\centering
	\includegraphics[width=\linewidth]{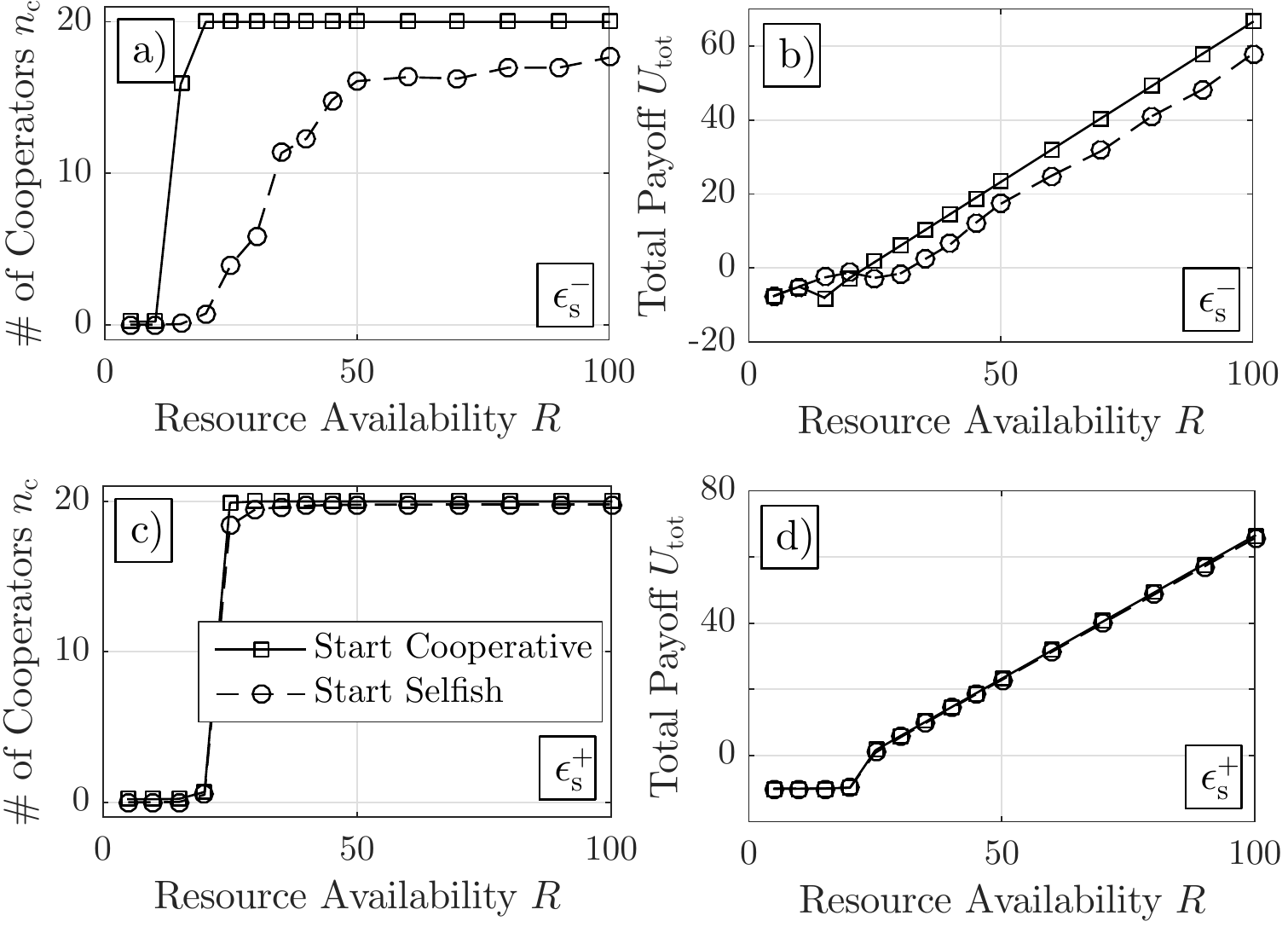}
	\caption{Mean number of cooperators and mean total payoff in a game where we vary the resource availability $\food$. In a) and b), the efficiency of the selfish players decreases as a function of the number of cooperative players. In c) and d), the efficiency of the selfish players increases as a function of the number of cooperative players.}
	\label{fig_game_vs_R}
\end{figure}

In Fig.~\ref{fig_game_vs_R}, we vary the resource availability $\food$. We observe that when $\food$ is sufficiently low, the players cannot benefit from cooperation. This is a common occurrence in biology when food is scarce; see \cite[Ch.~17]{Broom2013}. As $\food$ increases, the higher efficiency in cooperation overcomes its cost. Similar results are observed when using $\effSelfPos$ as when using $\effSelfNeg$; in the remainder of this work, we use $\effSelfNeg$ and assume that the selfish efficiency decreases with the efficiency of the cooperating players.

\begin{figure}[!tb]
\centering
\includegraphics[width=\linewidth]{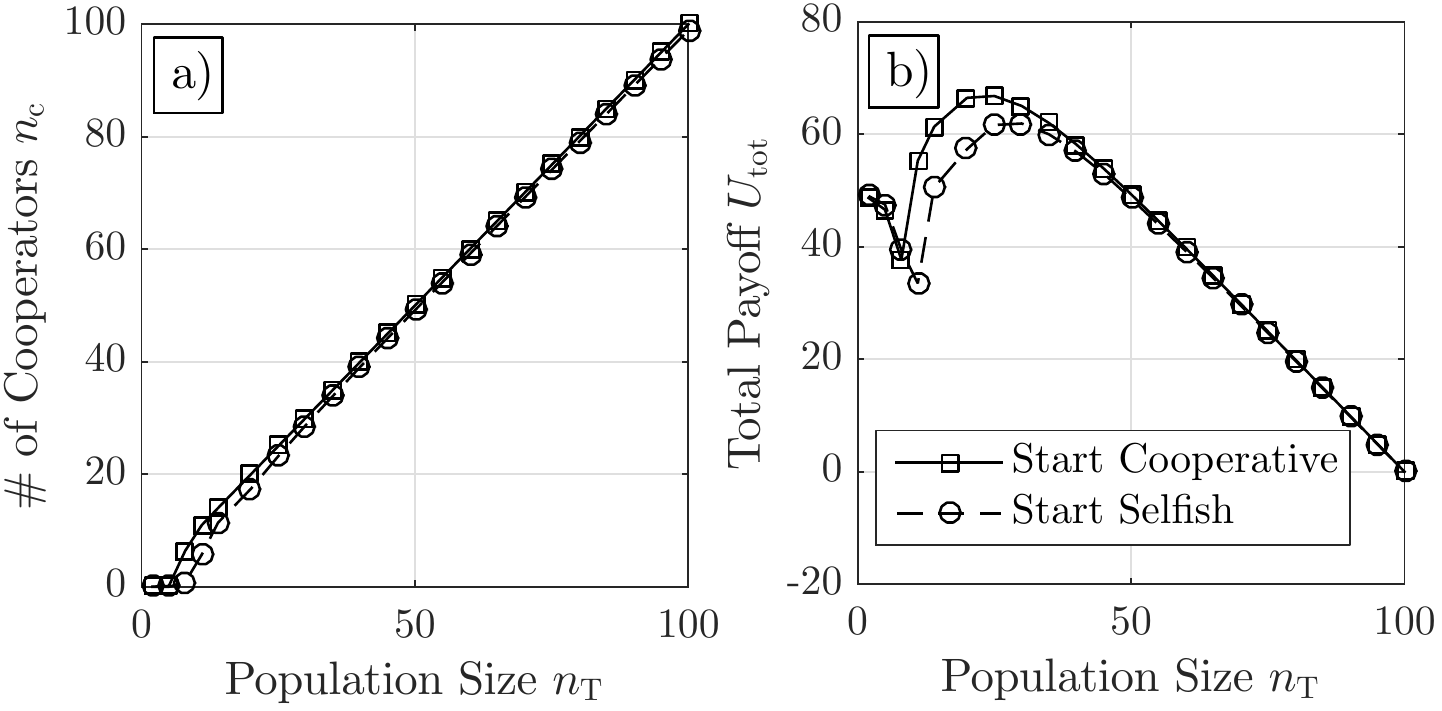}
\caption{a) Mean number of cooperators and b) mean total payoff in a game where we vary the total population $\numPop$.}
\label{fig_game_vs_pop}
\end{figure}

In Fig.~\ref{fig_game_vs_pop}, we vary the total population $\numPop$. When there are only a few players, they cannot benefit from cooperation and adding players reduces the total payoff $\payoffTotal$. However, consistent with quorum sensing, an increase in population size leads to cooperation and an increase in $\payoffTotal$. This trend continues until there are about 25 players, beyond which the diminishing increase in efficiency in (\ref{eqn_coop_efficiency}) cannot overcome the need to share the resource with more players. Even though we do not model population changes within an individual game, this observation supports the fact that bacteria populations saturate as they consume all of the resources in an environment, e.g., see \cite{Lambert2014}.

\begin{figure}[!tb]
\centering
\includegraphics[width=\linewidth]{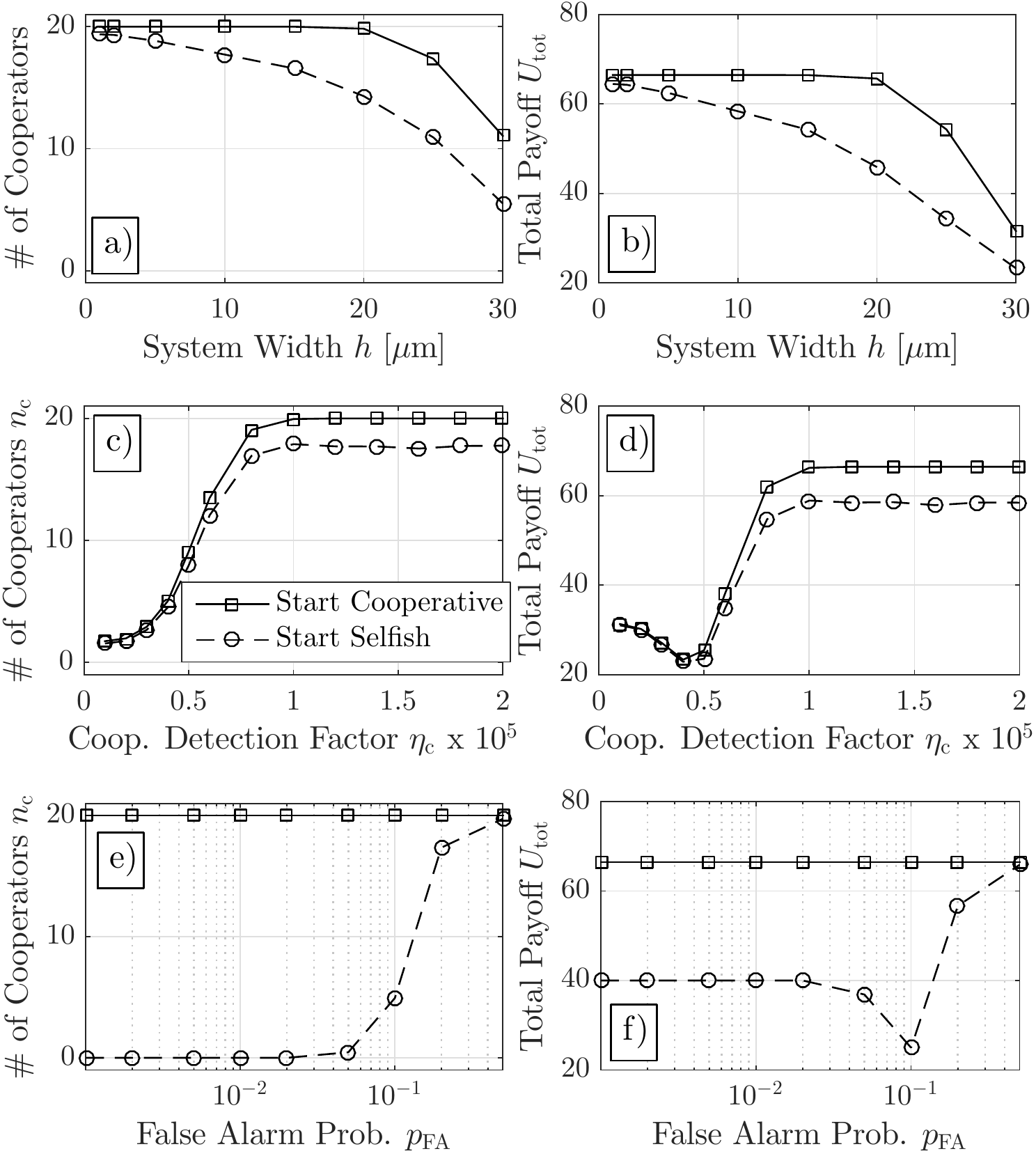}
\caption{a), c), e) Mean number of cooperators and b), d), f) mean total payoff in a game where we vary a), b) the width of the system $\boxWidth$, c), d) the cooperator detection factor $\detectMultCoop$, and e), f) the false alarm probability $\probFA$.}
\label{fig_game_vs_detection}
\end{figure}

In Fig.~\ref{fig_game_vs_detection}, we vary some of the detection parameters, i.e., the width $\boxWidth$ of the system, the cooperator detection factor $\detectMultCoop$, and the probability of false alarm $\probFA$. We observe that cooperation is less likely when other players are harder to detect, e.g., when the system increases in size or the cooperation detection factor decreases. This observation is consistent with the impact of population size in Fig.~\ref{fig_game_vs_pop} and the dynamics of quorum sensing. Even more interesting is the importance of having a non-negligible false alarm probability. As we see in Fig.~\ref{fig_game_vs_detection}e), the population needs uncertainty in classifying selfish players in order to be able to iterate from a fully selfish state to a cooperating one. This supports the claim that noise can replace explicit coordination in bacteria.

\section{Conclusions}
\label{sec_conclusions}

In this paper, we presented a simple payoff model for bacteria consuming a common resource and evaluated the model using game theory. We observed how cooperative behavior could be promoted or discouraged by the modification of system parameters, and that uncertainty in the local population can lead to higher payoffs than perfect knowledge. Future work could consider diverse populations with non-uniform payoff structures, more refined models to describe the payoffs and the estimation process, and the effects of evolution where successful strategies are preserved across generations of players.

\section*{Acknowledgment}

This work was supported in part by the Natural Sciences and Engineering Research Council of Canada (NSERC).

\bibliography{gt_refs}

\begin{thebibliography}{10}
\providecommand{\url}[1]{#1}
\csname url@samestyle\endcsname
\providecommand{\newblock}{\relax}
\providecommand{\bibinfo}[2]{#2}
\providecommand{\BIBentrySTDinterwordspacing}{\spaceskip=0pt\relax}
\providecommand{\BIBentryALTinterwordstretchfactor}{4}
\providecommand{\BIBentryALTinterwordspacing}{\spaceskip=\fontdimen2\font plus
\BIBentryALTinterwordstretchfactor\fontdimen3\font minus
  \fontdimen4\font\relax}
\providecommand{\BIBforeignlanguage}[2]{{%
\expandafter\ifx\csname l@#1\endcsname\relax
\typeout{** WARNING: IEEEtran.bst: No hyphenation pattern has been}%
\typeout{** loaded for the language `#1'. Using the pattern for}%
\typeout{** the default language instead.}%
\else
\language=\csname l@#1\endcsname
\fi
#2}}
\providecommand{\BIBdecl}{\relax}
\BIBdecl

\bibitem{Srivastava2005}
V.~Srivastava, J.~Neel, A.~Mackenzie, R.~Menon, L.~Dasilva, J.~Hicks, J.~Reed,
  and R.~Gilles, ``Using game theory to analyze wireless ad hoc networks,''
  \emph{IEEE Commun. Surv. Tutorials}, vol.~7, no.~4, pp. 46--56, 2005.

\bibitem{Schuster2008}
S.~Schuster, J.-U. Kreft, A.~Schroeter, and T.~Pfeiffer, ``Use of
  game-theoretical methods in biochemistry and biophysics.'' \emph{J. Biol.
  Phys.}, vol.~34, no. 1-2, pp. 1--17, Apr. 2008.

\bibitem{Berg2004}
H.~C. Berg, \emph{E. coli in Motion}.\hskip 1em plus 0.5em minus 0.4em\relax
  Springer, 2004.

\bibitem{Velicer2003}
G.~J. Velicer, ``Social strife in the microbial world,'' \emph{Trends
  Microbiol.}, vol.~11, no.~7, pp. 330--337, Jul. 2003.

\bibitem{Hummert2014}
S.~Hummert, K.~Bohl, D.~Basanta, A.~Deutsch, S.~Werner, G.~Thei{\ss}en,
  A.~Schroeter, and S.~Schuster, ``Evolutionary game theory: Cells as
  players,'' \emph{Mol. BioSyst.}, vol.~10, no.~12, pp. 3044--3065, Aug. 2014.

\bibitem{Atkinson2009}
S.~Atkinson and P.~Williams, ``Quorum sensing and social networking in the
  microbial world,'' \emph{J. R. Soc. Interface}, vol.~6, no.~40, pp. 959--978,
  Nov. 2009.

\bibitem{Lambert2014}
G.~Lambert, S.~Vyawahare, and R.~H. Austin, ``Bacteria and game theory: The
  rise and fall of cooperation in spatially heterogeneous environments,''
  \emph{Interface Focus}, vol.~4, no.~4, p. 20140029, Jun. 2014.

\bibitem{Travisano2004}
M.~Travisano and G.~J. Velicer, ``Strategies of microbial cheater control,''
  \emph{Trends Microbiol.}, vol.~12, no.~2, pp. 72--78, Feb. 2004.

\bibitem{Brown2001}
S.~P. Brown and R.~A. Johnstone, ``Cooperation in the dark: Signalling and
  collective action in quorum-sensing bacteria,'' \emph{Proc. R. Soc. London B
  Biol. Sci.}, vol. 268, no. 1470, pp. 961--965, May 2001.

\bibitem{Michelusi2017}
N.~Michelusi, ``On population density estimation via quorum sensing,'' in
  \emph{Proc. Can. Work. Inf. Theory}.\hskip 1em plus 0.5em minus 0.4em\relax
  IEEE, Jun. 2017, pp. 1--5.

\bibitem{Canzian2014}
L.~Canzian, K.~Zhao, G.~C.~L. Wong, and M.~van~der Schaar, ``A dynamic network
  formation model for understanding bacterial self-organization into
  micro-colonies,'' \emph{IEEE Trans. Mol. Biol. Multi-Scale Commun.}, vol.~1,
  no.~1, pp. 76--89, Mar. 2015.

\bibitem{Koca2017}
C.~Koca and O.~B. Akan, ``Anarchy vs. cooperation on internet of molecular
  things,'' \emph{IEEE Internet Things J.}, to appear.

\bibitem{Broom2013}
M.~Broom and J.~Rychtar, \emph{Game-Theoretical Models in Biology}.\hskip 1em
  plus 0.5em minus 0.4em\relax CRC Press, 2013.

\bibitem{Llatser2013}
I.~Llatser, A.~Cabellos-Aparicio, M.~Pierobon, and E.~Alarc{\'{o}}n,
  ``Detection techniques for diffusion-based molecular communication,''
  \emph{IEEE J. Sel. Areas Commun.}, vol.~31, no.~12, pp. 726--734, Dec. 2013.

\end{thebibliography}

\end{document}